\documentclass[iop]{emulateapj}
\usepackage{natbib}
\def\ie{{\it i.e.\ }}
\def\eg{{\it e.g.\ }}

\shorttitle{GSC 03144-595, a new triple-mode HADS}
\shortauthors{Mow et al.}

\begin{document}

\title{Rapid evolution of GSC 03144-595, a new triple-mode radially pulsating high-amplitude $\delta$-Scuti }
\author{
Benjamin Mow, Erik Reinhart, Samantha Nhim, 
and Richard Watkins
}
\affil{Department of Physics, Willamette University, Salem, OR 97301}
\email{rwatkins@willamette.edu}

\begin{abstract}

We present results of a multi-year study of the high-amplitude $\delta$ Scuti (HADS) star GSC 03144-595.   The star was observed between June and September in 2011 and 2014 for 13 nights and 28 nights respectively.   Based on our results we argue that GSC 03144-595 is a new triple-mode radially pulsating HADS, only the fifth discovered and only the second that has a fundamental frequency in the traditional $\delta$ Scuti regime.   While the frequencies and amplitudes of the fundamental and first harmonic were found to be unchanged between 2011 and 2014, we found that the amplitude of the second harmonic increased by  44\%, a form of evolution not previously seen.  This finding suggests that the second harmonic may be transient, thus explaining the scarcity of triple mode HADS stars.  

\end{abstract} 


\section{Introduction}  

Much has been learned about the solar interior from helioseismology, the study of small wave oscillations in the Sun (see, \eg, \citet{helio} and references therein).   This type of study is impossible to perform on other stars, as the stellar disks cannot be resolved.   However, the study of pulsating variable stars, known as astroseismology, can provide a similar window into the interiors of stars (see, \eg \citet{Percy}).   By applying Fourier analysis to measurements of the apparent magnitude vs. time of a pulsating variable star,  frequencies, amplitudes, and phases of its normal modes can be extracted.  These parameters can be compared to numerical models and used to refine models of stellar evolution.    

The advent of automated surveys searching for extrasolar planets (\eg {\it Kepler} and CoRoT), and gravitational microlensing (\eg EROS, MACHO, and OGLE) have also found an unprecedented number of pulsating variable stars.   This large influx of data has led to not only a much better understanding of the variety of pulsating variables, but also given us a richer view into the pulsations of individual stars.  The high quality of space-based data in particular has allowed us to observe much smaller amplitude pulsations than were previously possible, leading to exciting developments regarding small amplitude pulsators such as gamma Doradus ($\gamma$ Dor) stars and ``hybrid" stars, which show properties of both $\gamma$ Dor stars and delta Scuti ($\delta$ Sct) stars \citep{KeplerBradley}.   These stars are particularly interesting for astroseismology since their gravity modes probe much deeper regions of the stars than the pressure modes and mixed modes found in $\delta$ Sct stars.  

Even in the era of automated surveys, ground based observations of individual stars can still make important contributions, particularly in regard to studying nearby high-amplitude $\delta$ Sct (HADS) stars.  The magnitudes of these stars can be accurately measured from the ground, and their short periods mean that many periods can be observed over a typical observing season.  In addition, choices such as observation times, observation frequency, photometric filters used can be made much more flexibly in a observation campaign for an individual star.   Given that HADS stars are rare, estimated to make up only 0.3\% of the total Galactic population of $\delta$ Sct stars \citep{Lee08}, and that automated surveys still only cover a small portion of the sky, individual ground based observation campaigns to study HADS will remain important for years to come.  

The focus of this paper is the HADS star GSC 03144-595.   This star was reported by \citet{Pigulski09} to pulsate in two independent frequencies based on data from the All Sky Automated Survey \citep{ASAS}.   The star happens to lie in the {\it Kepler} field and both short cadence \citep{short} and long cadence \citep{long} observations are available in the Kepler Archive; it has the {\it Kepler} ID 6382916.   The star has an average V magnitude of 10.63, and {\it Kepler} observations have estimated that it has an effective temperature $T_{eff}= 6255\pm 200$ K, a surface gravity given by $\log g=4.159\pm0.5$, and an estimated radius of 1.498 $R_{\odot}$.   Like Cepheids, $\delta$ Sct stars have a period-luminosity relation; for example, \citet{PetChr} give the relation
\begin{equation}
M_V = -3.725\log P - 1.969,
\end{equation}
where $M_V$ is the absolute V-magnitude and $P$ is the period in days of the fundamental pulsation mode.   For GSC 03144-595 we have $P=0.2$ d, which implies that $M_V= 0.63$.   The distance modulus for this star is thus $\mu= m_V-M_V = 10.0$, suggesting the star is at a distance $d= 1,000$ pc.

\citet{Ulusoy13} detected a third independent frequency based on ground based observations taken in 2011 and observations taken by the {\it Kepler} satellite.  They also present ground based spectroscopic observations; by fitting H$\alpha$ and H$\beta$ lines they estimate that $T_{eff}= 6950\pm100 $K and $\log g = 3.7\pm0.1$.     It is interesting to note that this estimate of $T_{eff}$ disagrees with the {\it Kepler} value at a little greater than the 3$\sigma$ level.

																							Based on our own independent observations taken in 2011 and 2014, we will argue that all three independent modes found in GSC 03144-595 are radial modes, and that this star is a new triple-mode radially pulsating HADS star.   While multimode pulsation is more common for small amplitude $\delta$ Sct stars (see, for example, 	\citet{Piet13}, who found six triple mode small amplitude $\delta$ Sct stars in the OGLE-III data), it is relatively rare for HADS stars.  The AAVSO International Variable Star Index (VSX) \citep{VSX} lists only about eighty-five radial double-mode HADS (known as HADS(B) stars) out of almost 600 HADS stars.   Radial Triple mode HADS stars are particularly rare; there is not even a category for these stars in the VSX.   \citet{Wils08} lists only four known HADS stars that pulsate in three radial modes simultaneously.    Comparisons of single, double, and triple mode HADS stars may illuminate what determines the number of radial modes a star will pulsate in, something that is still not well understood.   The discovery of additional multi-mode radial pulsators would greatly help with these comparisons.    It would also be of interest to compare multi-mode HADS to double and triple mode RR Lyrae and Cepheid pulsators, many of which have been discovered in the Galactic Field, Galactic globular clusters, and in the Large and Small Magellanic clouds (see, \eg, \citet{Ogle08,Marqette09,Poleski13,Corwin08} and references therein).  
			
In Section~\ref{sec:obs} we describe our observations and give our results for the frequency, amplitudes, and phases of the three independent frequency modes as well as numerous combination modes.   Results are given for observations in both V and B bands for the years 2011 and 2014.   In Section~\ref{sec:modeid} we argue that the three independent frequency modes are radial.   We conclude in Section~\ref{sec:discussion}.

\section{Observations and Fourier Analysis}
\label{sec:obs}

GSC 03144-595 was observed for 13 nights (76 hrs.) between July and September of 2011, and again for 28 nights (142 hrs.) between June and September of 2014.    
All observations were carried out at Willamette University's observatory located 6.4 miles NW of Salem, Oregon, USA (lon 123.15$^\circ$ W, lat 45.02$^\circ$ N) , using an 11-inch aperture Celestron Edge HD telescope.     Alternating images were captured using V and B photometric filters with a QSI-532ws CCD camera.   Images were reduced using standard IRAF tasks including dark, bias, and flat frame corrections.   Differential magnitudes were calculated in IRAF using the VAPHOT task \citep{VAPHOT}.    There were no stars in our images that were comparable in both brightness and temperature to our variable.   We analyzed the V filter data using GSC 03144-00625 as a comparison star.   However, this star is significantly cooler than our variable and more than a magnitude dimmer in the B filter.   We found that the low flux of this comparison star introduced significant noise into our differential magnitudes.   To alleviate this problem, we used a different star, GSC 03144-00637, for a comparison star for our B filter data.   This star has about the same difference in magnitude from our variable in the B filter as GSC 03144-00625, but is brighter than the variable rather than dimmer.   The B filter results using either comparison star were very similar, but the use of GSC 03144-00637 resulted in significantly smaller uncertainties.  
Note that GSC 03144-00637 is also significantly cooler than our variable, so that in the V filter it is so much brighter than our variable that it is unsuitable for use as a comparison star.   

\begin{figure}
  \begin{center}
\includegraphics[trim=0.0cm 6cm 0cm 6cm, scale=0.48]{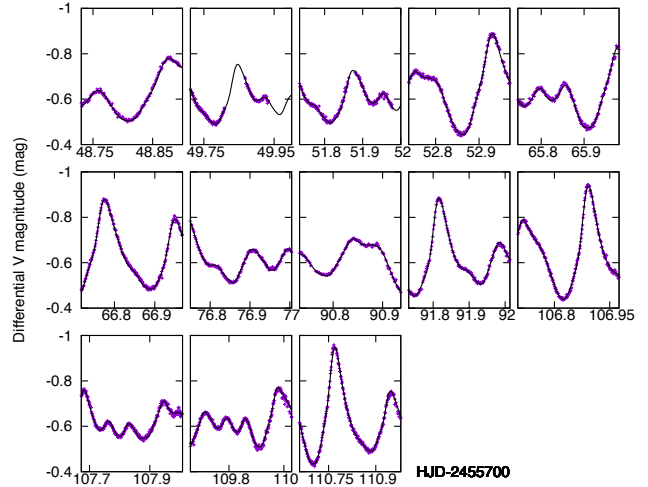}
 \end{center}
 \vspace{-0.5cm}
\caption{\small V filter light curves GSC 03144-595 with Fourier fit for 2011 data.
}
\label{fig:ts11}
\end{figure}

The Fourier analysis of the resulting light curves was carried out using PERIOD04 \citep{Period04}.  Three independent frequencies $f_1$, $f_2$, and $f_3$ were found together with many combination frequencies.   In Table~\ref{tab:freq} we show the frequencies, amplitudes, and phases found in both the 2011 and 2014 observations.   All of our frequencies are consistent with those found from the Kepler data by \citet{Ulusoy13}.  Only modes detected with signal to noise $S/N > 4.0$ are given \citep{Breger93}.   The frequencies given are for the V-filter data; the B-filter frequencies were consistent in all cases.   Only the three independent frequencies are determined in the fit and thus are given uncertainties.   Combination mode frequencies are determined from these frequencies.  We give phases for only the 2014 data, since phases are difficult to compare for data taken several years apart.   All uncertainties were calculated using Monte Carlo Simulations, as described in \citet{Period04}.  As a check on our results, we carried out our analysis on the two halves of our data independently.   We found that the frequencies and amplitudes for both halves were consistent with the results from the complete set.  

\begin{figure*}
  \begin{center}
\includegraphics[trim=0.0cm 6cm 0cm 6cm, scale=0.8]{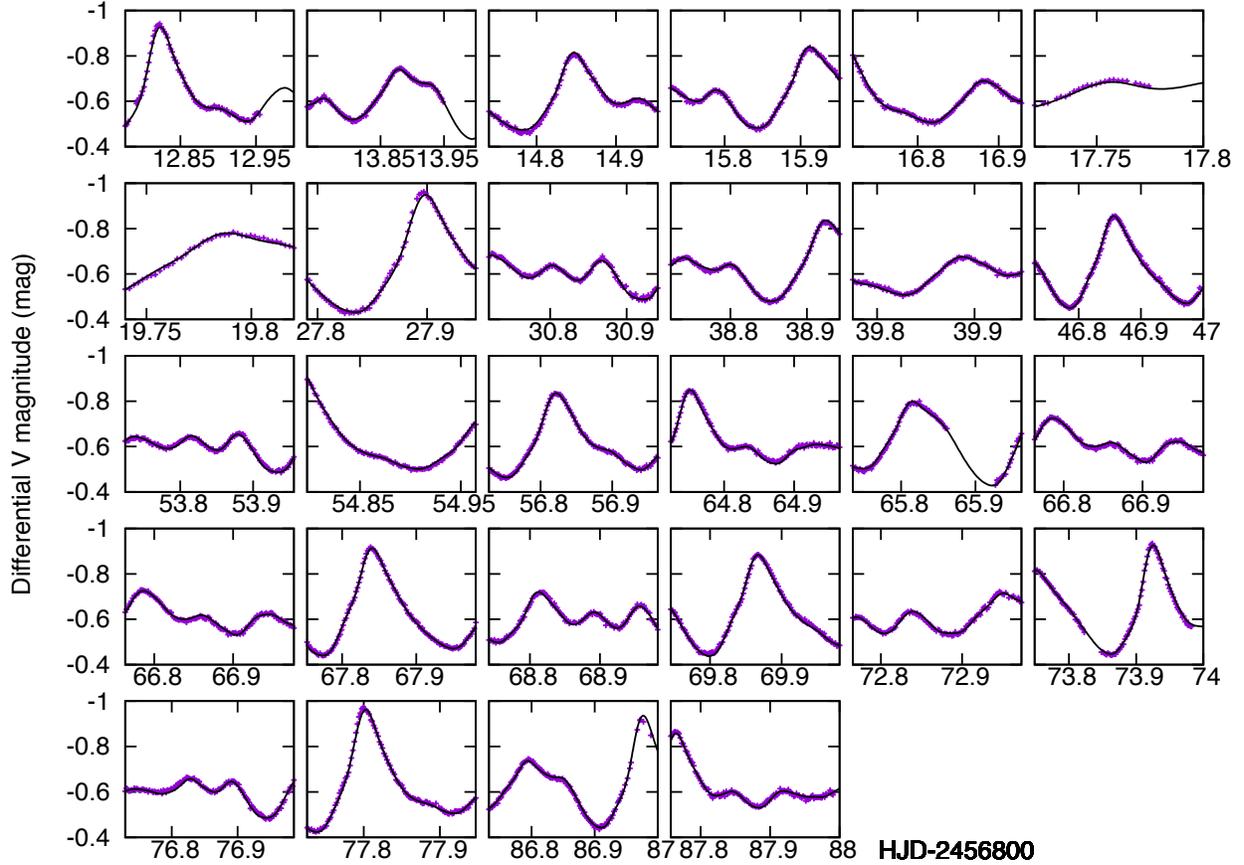}
 \end{center}
 \vspace{-0.5cm}
\caption{\small V filter light curves for GSC 03144-595 with Fourier fit for 2014 data.
}
\label{fig:ts14}
\end{figure*}

\begin{figure}
  \begin{center}
\includegraphics[trim=0.0cm 6cm 0cm 6cm, scale=0.39]{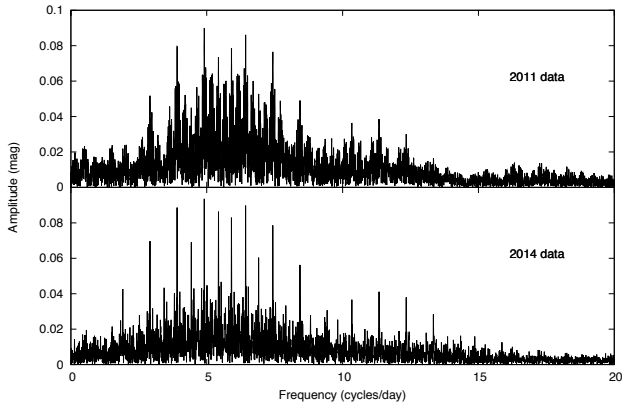}
 \end{center}
 \vspace{-0.5cm}
\caption{\small Fourier spectra for 2011 and 2014 data.
}
\label{fig:fourier}
\end{figure}

\begin{figure}
  \begin{center}
\includegraphics[trim=0.0cm 6cm 0cm 6cm, scale=0.39]{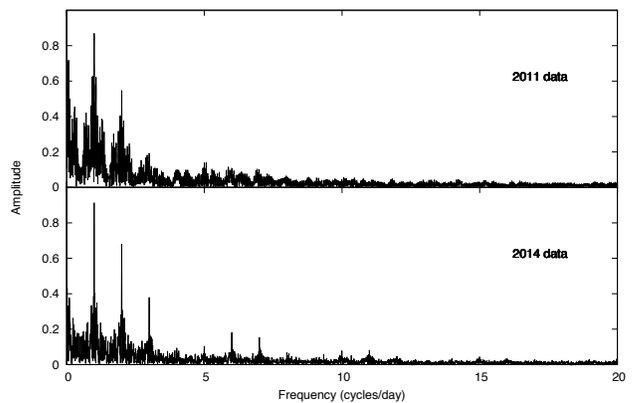}
 \end{center}
 \vspace{-0.5cm}
\caption{\small Spectral Window functions for 2011 and 2014 data.
}
\label{fig:window}
\end{figure}

In Fig.~\ref{fig:ts11} and Fig.~\ref{fig:ts14} we show light curves and Fourier fits for the 2011 and 2014 data respectively.   In Fig.~\ref{fig:fourier} we show the Fourier spectra for the 2011 and 2014 data.   Note that the highest peaks have the same frequency and amplitude in both datasets, but that the spectrum of the 2011 data is somewhat noisier due to having been calculated from many fewer nights of data.   In Fig.~\ref{fig:window} we show the window functions for the 2011 and 2014 data.  We see from Fig.~\ref{fig:window} that much of what appears to be noise in Fig.~\ref{fig:fourier} is actually coming from the window function.   Thus, in order to accurately analyze our data we used pre-whitening as described in \citet{Period04}.   

\begin{table*}
\caption{Frequencies, amplitudes, and phases of pulsation modes of GSC 03144-595 observed in both the V and B filters in the years 2011 and 2014.  Values in parentheses give the uncertainty in the final digit.  Values were given only for modes that were detected with signal to noise (S/N) ratio above 4.0; modes detected with S/N below 4.0 were left blank.    We give phases for only the 2014 data as phases measured several years apart are difficult to compare.  Combination modes in the bottom portion of the table contain the frequency $f_3$. Epoch of zero phase is HJD 2456800.}
\begin{tabular}{ccccccccc}
\hline\hline

 & f$^{11}$(d$^{-1}$) & f$^{14}$(d$^{-1}$) &$A_V^{11}$(mag) & $\ A_V^{14}$(mag)& $A_B^{11}$(mag)&$A_B^{14}$(mag) & $\Phi_V^{14}$(rad) & $\Phi_B^{14}$(rad) \\
\hline\hline
$f_1$ & 4.90990(2) & 4.90988(3)&0.0946(4) & 0.0951(4) & 0.1306(8) & 0.1339(5) & 4.081(4) & 4.136(4)\\
$f_2$ & 6.4319(2) & 6.43187(2)&0.0929(3) & 0.0929(4) & 0.1293(7) & 0.1294(8) & 1.806(5)  & 1.819(4)\\
\hline
$f_3$& 8.0351(2) & 8.03540(9)&0.0151(4) & 0.0215(5) & 0.0225(7) & 0.0303(6) & 4.05(2) & 3.97(2)\\
\hline
$f_1+f_2$ & 11.3418 & 11.3417&0.0358(4) & 0.0364(4) & 0.0486(8) & 0.0492(6) & 2.43(1) & 2.44(1)\\
$f_2-f_1$ & 1.5219 & 1.5220 & 0.0199(4) & 0.0185(4) & 0.0280(7) & 0.0275(6) & 2.44(2) & 2.42(2)\\
$2f_1$ &9.8198 &  9.8197& 0.0197(4) & 0.0201(3) & 0.0270(7) & 0.0265(6) & 4.69(2) & 4.71(2)\\
$2f_2$ & 12.8638 & 12.8638 &0.0148(4) & 0.0150(4) & 0.0192(8) & 0.0193(6) & 4.54(3) & 4.51(3)\\
$2f_1+f_2$ & 16.2517 & 16.2516& 0.0117(4) & 0.0120(4) & 0.0161(7) & 0.0155(7) & 1.47(3)& 1.51(4)\\
$2f_2-f_1$ & 7.9539 & 7.9539&0.0090(4) & 0.0081(4) & 0.0116(8) & 0.0104(7) & 0.15(4) & 0.10(6)\\
$2f_2+f_1$ & 17.7737 & 17.7736&0.0064(4) & 0.0062(4) & 0.0096(9) & 0.0095(7) & 6.19(6)  & 6.21(7)\\
$2f_1-f_2$ & 3.3879 & 3.3878 & 0.0061(8) & 0.0058(4) & 0.0115(9) & 0.0113(6) & 0.25(7) & 0.61(6) \\
$2f_1+2f_2$ &22.6836 & 22.6835 & 0.0058(4) & 0.0053(3) & 0.0071(7) & 0.0071(6) & 5.72(7) & 5.70(9) \\
$f_1+3f_2$ & 24.2056 & 24.2055 & 0.0036(4) & 0.0036(4) & 0.0045(8) & 0.0051(7) & 3.2(1)& 3.2(1) \\
$3f_1+ 2f_2$ & 27.5932 & 27.5933 & 0.0033(4) & 0.0033(4) & 0.0050(7) & 0.0046(6) & 6.1(1) & 6.2(1) \\
$3f_1$ & 14.7297 & 14.7295 & 0.0024(4) & 0.0031(4) & 0.0035(8) & 0.0050(6) & 5.3(1) & 5.4(1) \\
 $3f_1+3f_2$ & 34.0254 & 34.0252 & 0.0019(3) & 0.0020(4) & 0.0024(8) & 0.0028(6) & 4.1(2) & 4.2(2) \\
  $3f_1+f_2$ & 21.1616 & 21.1614 & 0.0027(4) & 0.0025(4) & 0.0034(7) & 0.0032(6) & 2.0(1) & 2.0(2) \\
 $2f_1+3f_2$ & 29.1155 & 29.1153 & 0.0020(4) & 0.0024(3) & 0.0027(8) & 0.0027(6) & 3.6(2) & 3.8(2) \\
  $3f_2$ & 19.2957 &19.2956 & 0.0032(3) & 0.0029(4) &0.0033(7) &0.0032(7) & 2.2(1) & 2.3(2) \\
  $3f_2-f_1$ & 14.3858 &14.3858 & 0.0018(4) & 0.0015(4) &0.0014(9) & & 4.1(3) &  \\
   $3f_1-f_2$ & 8.2978 & 8.2977 & 0.0018(4) & 0.0012(4) & & & 0.1(3) &  \\
  $2f_1+4f_2$ & 35.5474 &35.5472 &  & 0.0012(3) && & 1.6(3) &  \\
\hline
$f_2+f_3$ & 14.4669 & 14.4673 & 0.0045(4) & 0.0058(4) & 0.0049(7) & 0.0077(6) & 0.90(8) & 0.87(8) \\
$f_1+f_3$ & 12.9450 & 12.9452 & 0.0040(3) & 0.0061(4) & 0.0057(8) & 0.0082(6) & 4.85(6) & 4.74(7) \\
$f_3-f_2$ &1.6032 & 1.6035 & 0.0034(4) & 0.0043(4) & & & 0.42(9) &  \\
$f_2-f_1+f_3$ &9.5571 & 9.5574 & 0.0030(4) & 0.0031(4) & 0.0040(8) & 0.0046(6)& 1.9(1) & 1.8(1) \\
$f_3-f_1$ & 3.1252 & 3.1255& 0.0042(4) & 0.0041(4) & & & 3.58(9)& \\
$f_1+f_2-f_3$ & 3.3067 & 3.3063 & 0.0032(4) & 0.0042(4) &  &  & 3.0(1)&  \\
$2f_1+f_3$ & 17.8549 & 17.8551 & 0.0019(4) & 0.0026(4) & 0.0030(8) & 0.0030(5) & 4.9(2) & 4.9(2) \\
 $f_1+2f_2+f_3$ & 25.8088 & 25.8090&  0.0017(4) & 0.0016(4) & &0.0023(6) & 5.4(2) & 5.1(3)\\
 $f_1-f_2+f_3$ &6.5131  & 6.5134& 0.0015(3) & 0.0026(4) &  & 0.0040(6) & 0.5(1) & 0.4(2)  \\
  $2f_2+f_3$ & 20.8989 & 20.8992& 0.0013(4) & 0.0024(4) & &0.0031(6) & 4.8(2) & 5.0(2) \\
  $f_1+f_2+f_3$ & 19.3769 & 19.3771& 0.0023(4) & 0.0024(3) & 0.0035(7)&0.0033(7) & 2.4(1)& 2.2(2) \\
  $2f_2-f_1+f_3$ & 15.9890 & 15.9893& 0.0017(4) & 0.0017(4) & & & 6.2(2) & \\
  $2f_2-f_3$ &  & 4.8284 & & 0.0014(4) & & & 0.0(3) & \\
  $2f_1+2f_2+f_3$ &  &30.7189 &  & 0.0014(4) && & 5.6(2) &  \\
 $2f_1+f_2+f_3$ &  &24.2870 &  & 0.0014(4) && 0.0020(6)& 2.1(3)& 1.9(4) \\
$2f_1+f_2-f_3$ & &8.2162 & & 0.0014(4) &&&3.6(2) &  \\

\label{tab:freq}
\end{tabular}
\end{table*}

\section{Mode Identification}
\label{sec:modeid}

In Table~\ref{tab:freq} we see consistency  in both frequency and amplitude over the time interval for the first two independent modes.   However, although the third independent mode has remained constant in frequency, it's amplitude has increased by 44\% between 2011 and 2014.   This increase corresponds to almost 12 standard deviations, and thus cannot be due to random error.   Further confirmation of this increase comes from examining the combination modes.   While combinations of the first two fundamental frequencies are statistically unchanged, more than half of the combination modes involving $f_3$ show significant increases in amplitude.   The factor by which these combination mode amplitudes have increased is comparable to the factor by which the $f_3$ mode has increased.   This result makes sense since the combination modes involving $f_3$ are generated by nonlinear interactions between the third mode and the $f_1$ and $f_2$ modes.  Thus we would expect the amplitude of the combination modes involving $f_3$ to be proportional to the amplitude of the $f_3$ mode.

\citet{Ulusoy13} has reported multi-site observations of GSC 03144-595 from 2011.   They argue that the the lowest two independent frequency modes $f_1$ and $f_2$ are non-radial, $l=1$ modes, and that the identification of the third mode is uncertain due to its small amplitude.   Here we argue that all three of the independent pulsation modes are radial, and that GSC 03144-595 is a new triple-mode radial pulsator.

We first consider the period ratios for the independent modes.    \citet{Ulusoy13} claim that the observed period ratio $P_1/P_0=0.763$ of GSC 03144-595 differs significantly from the expected value for radial first overtone and fundamental modes.  However, there is no accepted lower bound for this period ratio, and several double mode pulsators have been observed to have a similar ratio \citep{Poretti05}.   In fact, this period ratio is very close to those found for triple-mode radial pulsators \citep{Wils08}, which tend to have smaller values of $P_1/P_0$ than double-mode pulsators.   

\begin{table*}
\caption{Amplitude ratios and phase differences for the three independent pulsation modes of GSC 03144-595 as determined from our 2011 data, our 2014 data, and the observations made by  \protect\citet{Ulusoy13} in 2011.  For the \protect\citet{Ulusoy13} results the first uncertainty is the one reported in the paper and the second uncertainty, in parenthesis, is calculated from their data using Monte Carlo methods.}
\begin{tabular}{cccc||ccc}
\hline\hline
 & $A_V^{11}/A^{11}_B$ & $A_V^{14}/A^{14}_B$ & $A_V^{U}/A^{U}_B$ & $\Phi^{11}_V-\Phi^{11}_B $ & $\Phi^{14}_V-\Phi^{14}_B$ & $\Phi^{U}_V-\Phi^{U}_B$\\
\hline\hline
$f_1$ & 0.725(5) & 0.710(4) & 0.725$\pm$0.004($\pm$0.006) &-0.055(7) & -0.055(5) & 0.07$\pm$0.02($\pm$0.01) \\
$f_2$ & 0.719(5) & 0.718(5) & 0.732$\pm$0.004($\pm$0.005) &-0.030(7) & -0.008(6) & -0.04$\pm$0.02($\pm$0.01) \\
$f_3$ & 0.67(3) & 0.71(2) & 0.97$\pm$0.04($\pm$0.2) & -0.01(5) & 0.08(3) & 0.4$\pm$0.1($\pm$0.6) \\
\label{tab:ratio}
\end{tabular}
\end{table*}

Of even more interest is the ratio of the periods of the second and first harmonics, $P_2/P_1$.   This ratio is observed to be in a narrow range around 0.8 in multimode pulsators (see \eg Fig. 5 in \citet{triplemode06}).  Given the much narrower range of observed values, this ratio is in some ways more diagnostic than $P_1/P_0$.   The ratio for GSC 03144-595
is $P_2/P_1= 0.800$, within the observed range for general multi-mode pulsators and very similar to the values found in triple mode $\delta$ Scutis  \citep{Wils08}.   In fact, the four known triple mode pulsators in the Galaxy \citep{Wils08} have an average value $\langle P_2/P_1\rangle = 0.801$ with a standard deviation of only $\sigma_{P_2/P_1} = 0.001$.   The fact that the period ratios of GSC 03144-595 correspond so well to those of known triple-mode radial pulsators is strongly suggestive that it is in fact a new addition to this class.  Finally, the ratio of the V-band amplitudes of the first harmonic and the fundamental for GSC 03144-595 is 1.02, a result that matches that of the two triple mode $\delta$ Scutis with fundamental frequencies closest to that of GSC 03144-595, which have ratios of 1.04 and 1.00.

The main evidence that \citet{Ulusoy13} give against the independent modes of GSC 03144-595 being radial is an analysis of amplitude ratios and phase differences in different wavebands using multicolor photometry.  Changes in brightness of pulsating variable stars arise from both changes in surface area and changes in temperature, typically acting in opposition.   Ratios of mode amplitudes measured in different filters can indicate the degree of temperature change associated with a pulsation mode; for example, an amplitude ratio for two filters near unity would indicate that the pulsation mode has little effect on the surface temperature of the star.   Since different types of pulsation modes, \eg modes with different $l$ values, effect the temperature differently, we would expect amplitude ratios to be indicative of mode type.  Similarly, phase differences indicate the relative timing of temperature and area changes and should also be diagnostic.  

\begin{table}
\caption{Amplitude ratios for the three independent pulsation modes of V823 Cassiopeia and GSC 762-110 determined from results reported in \protect\citet{triplemode06} and \protect\citet{Wils08} respectively.}
\begin{tabular}{cccc}
\hline\hline
& freq.(d$^{-1})$ & $A_V/A_B$ &  $\Phi_V-\Phi_B $ \\
\hline\hline
V823 Cass &$f_1$ & 0.742(4) & -0.039(2) \\
&$f_2$ & 0.738(2) & -0.017(2) \\
&$f_3$ & 0.76(2) & -0.01(2) \\
\hline\hline
GSC 762-110&$f_1$ & 0.778(4) & -0.007(5) \\
&$f_2$ & 0.790(5) &  0.010(6)\\
&$f_3$ & 0.81(1) & 0.02(1)\\
\hline\hline

\label{tab:triple}
\end{tabular}
\end{table}

\citet{Ulusoy13} compare their observational results to theoretical models.   Their main results are given in Fig. 5 and Table 5 of their paper, where they compare their measured amplitude ratios and phase differences to model results for different $l$ values as a function of frequency.   Based on this figure, they claim that of the three independent modes, those corresponding to $f_1$ and $f_2$ are $l=1$ modes and that the $f_3$ mode could not be reliably identified due to its low amplitude.   The authors claim that the radial mode is ruled out for $f_1$ and $f_2$ but do not attempt to quantify the statistical certainty of this claim.   However, an examination of the phase difference plot shows that, with the possible exception of the $f_3$ mode, which has large error bars, the observations do not agree in a statistical sense with any of the models.   In particular, for the $f_1$ and $f_2$ modes there is very little phase difference between the different wavebands, while the models show the phase difference increasing with increasing wavelength for $l\ge 1$ and decreasing with increasing wavelength for $l=0$.   The failure of the models to match the observed phase differences suggests a problem with this analysis, and calls into question its conclusions.

While the data we have collected is only in the V and B bands, it is still interesting to examine our results for the amplitude ratios $A_V/A_B$ and phase differences $\Phi_V-\Phi_B$ and compare them to those of \citet{Ulusoy13}.      
In Table~\ref{tab:ratio} we show amplitude ratios$A_V/A_B$ and phase difference $\Phi_V-\Phi_B$ for the three independent modes in both our 2011 and 2014 data.   For comparison, we also show the amplitudes and phase differences for these modes reported by \citet{Ulusoy13}.

Our results from Table~\ref{tab:ratio} show the same values, within statistical uncertainties, for the amplitude ratio for all three independent modes.  These values are also consistent across both observation years, even after the evolution of the $f_3$ mode.    Given the relatively small uncertainties, this strongly suggests that all three independent modes are of the same type, \ie have the same $l$ value.   This result, taken together with our discussion of the period ratios above, suggests that all three of the independent modes are radial, and that they are in fact the fundamental, the first harmonic, and the second harmonic.   

While the phase differences $\Phi_V-\Phi_B$ are consistent across the two study years, they are significantly different between the modes.   In order to understand this difference,  it is useful to compare our results to known triple-mode radially pulsating HADS stars.   In Table~\ref{tab:triple} we show amplitude ratios and phase differences for the fundamental ($f_1$), first harmonic ($f_2$), and second harmonic ($f_3$) of V823 Cassiopeia \citep{triplemode06} and GSC 762-110 \citet{Wils08}.   We see that the amplitude ratio values are comparable, although both triple mode stars seem to have slightly larger ratios, with GSC 762-110 having the largest.   As in GSC 03144-595, the amplitude ratios are consistent between the modes.   However, we see that the phase difference does not appear to be the same between the modes, but rather is most negative for $f_1$, less negative for $f_2$, and even less negative, or slightly positive, for $f_3$   This is particularly evident in V823 Cassiopeia, where the uncertainties are small enough that this pattern is quite significant.   This is the same pattern seen in GSC 03144-595 above, again showing a similarity between it and the triple mode HADS.   Overall, this provides strong evidence that while amplitude ratio appears to be the same for all radial modes, the phase difference has an additional dependence on what particular mode is being considered, with lower frequency modes having more negative phase differences.

Table~\ref{tab:ratio} also shows that, given the reported uncertainties, our results for the phase difference for $f_1$ and the amplitude ratio and phase difference for $f_3$ are in significant disagreement with those reported by  \citet{Ulusoy13}.    However,  \citet{Ulusoy13} used analytical methods \citep{Breger99} to estimate their uncertainties from their least squares fit, a method which assumes an ideal case.   We have applied the more reliable method of Monte Carlo Simulation to their data to provide alternative estimates of the uncertainties, which are given in Table~\ref{tab:ratio} after the reported uncertainties in parenthesis.   While the uncertainties are relatively unchanged for the $f_1$ and $f_2$ results, the Monte Carlo uncertainties for the $f_3$ mode are much larger for both the amplitude ratio and the phase difference.  The increased uncertainties primarily  arise from much larger Monte Carlo uncertainties in the results from the B filter data, which appears to be noisier than the other filters.  This is due in part to the fact that \citet{Ulusoy13} used GSC 03144-00625 as a comparison star for the B filter data as well as the V filter data; as discussed above, the dimness of this star in the B filter makes it a less than ideal comparison star.   

Given the uncertainties calculated using Monte Carlo simulations, we see almost total agreement between our results and those of \citet{Ulusoy13}.   The only exception is the phase difference of the $f_1$ mode.   
Note that although the amplitude ratios for $f_2$ might appear to disagree, the uncertainty of the difference between our results and the Ulusoy results is given by $\sigma_{diff}= \sqrt{(0.05)^2 + (0.05)^2}= 0.071$.  Given this uncertainty, our amplitude ratios differ from those reported by \citet{Ulusoy13} at only about the 2$\sigma$ level.   

\section{Discussion}
\label{sec:discussion}

We have argued that GSC 03144-595 is a new triple-mode radially pulsating HADS.   Using data that we have collected over two different years, we have shown that this star resembles known triple-mode radially pulsating HADS stars in its period ratios $P_0/P_1$ and $P_1/P_2$, its amplitude ratios $A_V/A_B$, and its phase differences $\Phi_V-\Phi_B$.   These results are consistent between the two years of data.

Taken together, our examination of amplitude ratios and phase differences in triple-mode radially pulsating HADS has revealed an interesting pattern.   While amplitude ratios for radial modes appear to be the same for all radial modes, phase differences exhibit a characteristic pattern where they are increasingly negative with decreasing frequency.   This pattern has not been previously noted in the literature.   

Having two years of data separated by two intervening years allowed us to look for evolution of GSC 03144-595.   We found no significant changes in the frequencies of pulsation; however, while the amplitudes of the fundamental and first overtone were consistent in the two datasets, the amplitude of the second overtone increased by 44\%, corresponding to a change of almost 12 standard deviations.     This is a type of evolution not previously reported on.  The rapid change of the amplitude of the second overtone in GSC 03144-595 suggests that this mode may be transitory.   If indeed second overtones are transitory, this fact could explain why triple mode pulsators are rare; if second overtone modes appear and disappear over short timescales, then observing them may require making observations of a star in a specific time window.   

The discovery that GSC 03144-595 is a triple-mode radial pulsator brings the number of known triple-mode HADS to five.   More importantly, GSC 03144-595 is only the second known triple mode HADS, together with GSC 762-110, whose fundamental frequency is high enough to fall in the traditional $\delta$ Scuti regime.   Although the other three have frequencies more typical of RR Lyrae, they are considered to be evolved HADS stars \citep{Wils08}.  

Our results suggest that multi-mode HADS stars may evolve much faster than previously thought.   Frequent monitoring of these stars will help us better understand the dynamics of this fascinating class of variables.

\acknowledgements

We thank Rick Baumann, a Willamette University Alum, for generous financial support which made this research possible.   We also thank Willamette University's SCRP and Carson Grant programs for additional financial support.    We thank Richard Berry for inspiring us to study variable stars and for providing valuable technical advice.   

\bibliography{var}

\begin{thebibliography}{}
\expandafter\ifx\csname natexlab\endcsname\relax\def\natexlab#1{#1}\fi

\bibitem[{{Bradley} {et~al.}(2015){Bradley}, {Guzik}, {Miles}, {Uytterhoeven},
  {Jackiewicz}, \& {Kinemuchi}}]{KeplerBradley}
{Bradley}, P.~A., {Guzik}, J.~A., {Miles}, L.~F., {et~al.} 2015, \aj, 149, 68

\bibitem[{{Breger} {et~al.}(1993){Breger}, {Stich}, {Garrido}, {Martin},
  {Jiang}, {Li}, {Hube}, {Ostermann}, {Paparo}, \& {Scheck}}]{Breger93}
{Breger}, M., {Stich}, J., {Garrido}, R., {et~al.} 1993, \aap, 271, 482

\bibitem[{{Breger} {et~al.}(1999){Breger}, {Handler}, {Garrido}, {Audard},
  {Zima}, {Papar{\'o}}, {Beichbuchner}, {Li}, {Jiang}, {Liu}, {Zhou}, {Pikall},
  {Stankov}, {Guzik}, {Sperl}, {Krzesinski}, {Ogloza}, {Pajdosz}, {Zola},
  {Thomassen}, {Solheim}, {Serkowitsch}, {Reegen}, {Rumpf}, {Schmalwieser}, \&
  {Montgomery}}]{Breger99}
{Breger}, M., {Handler}, G., {Garrido}, R., {et~al.} 1999, \aap, 349, 225

\bibitem[{{Corwin} {et~al.}(2008){Corwin}, {Borissova}, {Stetson}, {Catelan},
  {Smith}, {Kurtev}, \& {Stephens}}]{Corwin08}
{Corwin}, T.~M., {Borissova}, J., {Stetson}, P.~B., {et~al.} 2008, \aj, 135,
  1459

\bibitem[{{Deeg} \& {Doyle}(2001)}]{VAPHOT}
{Deeg}, H.~J., \& {Doyle}, L.~R. 2001, in Third Workshop on Photometry, ed.
  W.~J. {Borucki} \& L.~E. {Lasher}, 85

\bibitem[{{Gilliland} {et~al.}(2010){Gilliland}, {Jenkins}, {Borucki},
  {Bryson}, {Caldwell}, {Clarke}, {Dotson}, {Haas}, {Hall}, {Klaus}, {Koch},
  {McCauliff}, {Quintana}, {Twicken}, \& {van Cleve}}]{short}
{Gilliland}, R.~L., {Jenkins}, J.~M., {Borucki}, W.~J., {et~al.} 2010, \apjl,
  713, L160

\bibitem[{{Jenkins} {et~al.}(2010){Jenkins}, {Caldwell}, {Chandrasekaran},
  {Twicken}, {Bryson}, {Quintana}, {Clarke}, {Li}, {Allen}, {Tenenbaum}, {Wu},
  {Klaus}, {Van Cleve}, {Dotson}, {Haas}, {Gilliland}, {Koch}, \&
  {Borucki}}]{long}
{Jenkins}, J.~M., {Caldwell}, D.~A., {Chandrasekaran}, H., {et~al.} 2010,
  \apjl, 713, L120

\bibitem[{{Jurcsik} {et~al.}(2006){Jurcsik}, {Szeidl}, {V{\'a}radi}, {Henden},
  {Hurta}, {Lakatos}, {Posztob{\'a}nyi}, {Klagyivik}, \&
  {S{\'o}dor}}]{triplemode06}
{Jurcsik}, J., {Szeidl}, B., {V{\'a}radi}, M., {et~al.} 2006, \aap, 445, 617

\bibitem[{{Lee} {et~al.}(2008){Lee}, {Kim}, {Shin}, {Lee}, \& {Jin}}]{Lee08}
{Lee}, Y.-H., {Kim}, S.~S., {Shin}, J., {Lee}, J., \& {Jin}, H. 2008, \pasj,
  60, 551

\bibitem[{{Lenz} \& {Breger}(2005)}]{Period04}
{Lenz}, P., \& {Breger}, M. 2005, Communications in Asteroseismology, 146, 53

\bibitem[{{Marquette} {et~al.}(2009){Marquette}, {Beaulieu}, {Buchler},
  {Szab{\'o}}, {Tisserand}, {Belghith}, {Fouqu{\'e}}, {Lesquoy}, {Milsztajn},
  {Schwarzenberg-Czerny}, {Afonso}, {Albert}, {Andersen}, {Ansari}, {Aubourg},
  {Bareyre}, {Charlot}, {Coutures}, {Ferlet}, {Glicenstein}, {Goldman},
  {Gould}, {Graff}, {Gros}, {Ha{\"i}ssinski}, {Hamadache}, {de Kat}, {Le
  Guillou}, {Loup}, {Magneville}, {Maurice}, {Maury}, {Moniez},
  {Palanque-Delabrouille}, {Perdereau}, {Rahal}, {Rich}, {Spiro}, \&
  {Vidal-Madjar}}]{Marqette09}
{Marquette}, J.~B., {Beaulieu}, J.~P., {Buchler}, J.~R., {et~al.} 2009, \aap,
  495, 249

\bibitem[{Percy(2007)}]{Percy}
Percy, J. 2007, {Understanding Variable Stars} (Cambridge University Press)

\bibitem[{{Petersen} \& {Christensen-Dalsgaard}(1999)}]{PetChr}
{Petersen}, J.~O., \& {Christensen-Dalsgaard}, J. 1999, \aap, 352, 547

\bibitem[{{Pietrukowicz} {et~al.}(2013){Pietrukowicz}, {Dziembowski},
  {Mr{\'o}z}, {Soszy{\'n}ski}, {Udalski}, {Poleski}, {Szyma{\'n}ski}, {Kubiak},
  {Pietrzy{\'n}ski}, {Wyrzykowski}, {Ulaczyk}, {Koz{\l}owski}, \&
  {Skowron}}]{Piet13}
{Pietrukowicz}, P., {Dziembowski}, W.~A., {Mr{\'o}z}, P., {et~al.} 2013,
  \actaa, 63, 379

\bibitem[{{Pigulski} {et~al.}(2009){Pigulski}, {Pojma{\'n}ski}, {Pilecki}, \&
  {Szczygie{\l}}}]{Pigulski09}
{Pigulski}, A., {Pojma{\'n}ski}, G., {Pilecki}, B., \& {Szczygie{\l}}, D.~M.
  2009, \actaa, 59, 33

\bibitem[{{Pojma{\'n}ski}(2004)}]{ASAS}
{Pojma{\'n}ski}, G. 2004, Astronomische Nachrichten, 325, 553

\bibitem[{{Poleski}(2013)}]{Poleski13}
{Poleski}, R. 2013, \apj, 778, 147

\bibitem[{{Poretti} {et~al.}(2005){Poretti}, {Su{\'a}rez}, {Niarchos},
  {Gazeas}, {Manimanis}, {van Cauteren}, {Lampens}, {Wils}, {Alonso}, {Amado},
  {Belmonte}, {Butterworth}, {Martignoni}, {Mart{\'{\i}}n-Ruiz}, {Moskalik}, \&
  {Robertson}}]{Poretti05}
{Poretti}, E., {Su{\'a}rez}, J.~C., {Niarchos}, P.~G., {et~al.} 2005, \aap,
  440, 1097

\bibitem[{{Soszynski} {et~al.}(2008){Soszynski}, {Poleski}, {Udalski},
  {Szymanski}, {Kubiak}, {Pietrzynski}, {Wyrzykowski}, {Szewczyk}, \&
  {Ulaczyk}}]{Ogle08}
{Soszynski}, I., {Poleski}, R., {Udalski}, A., {et~al.} 2008, \actaa, 58, 163

\bibitem[{{Toomre} \& {Thompson}(2015)}]{helio}
{Toomre}, J., \& {Thompson}, M.~J. 2015, \ssr, doi:10.1007/s11214-015-0147-x

\bibitem[{{Ulusoy} {et~al.}(2013){Ulusoy}, {Ula{\c s}}, {G{\"u}lmez}, {Balona},
  {Stateva}, {Iliev}, {Dimitrov}, {Kobulnicky}, {Pickering}, {Fox Machado},
  {{\'A}lvarez}, {Michel}, {Antoniuk}, {Shakhovskoy}, {Pit}, {Damasso},
  {Cenadelli}, \& {Carbognani}}]{Ulusoy13}
{Ulusoy}, C., {Ula{\c s}}, B., {G{\"u}lmez}, T., {et~al.} 2013, \mnras, 433,
  394

\bibitem[{{Watson} {et~al.}(2015){Watson}, {Henden}, \& {Price}}]{VSX}
{Watson}, C., {Henden}, A.~A., \& {Price}, A. 2015, VizieR Online Data Catalog,
  1

\bibitem[{{Wils} {et~al.}(2008){Wils}, {Rozakis}, {Kleidis}, {Hambsch}, \&
  {Bernhard}}]{Wils08}
{Wils}, P., {Rozakis}, I., {Kleidis}, S., {Hambsch}, F.-J., \& {Bernhard}, K.
  2008, \aap, 478, 865

\end{thebibliography}
\end{document}